\documentstyle[preprint,aps]{revtex}

\begin{document}

\draft

\title{Spatial Variation of Correlation Times for 1D Phase Turbulence}

\author{David~A.\ Egolf}
\address{
Department of Physics and the Center for Nonlinear and Complex Systems\\
Duke University, Durham, NC 27708-0305; dae@phy.duke.edu
}

\author{Henry~S.\ Greenside}
\address{Department of Computer Science, Department of
Physics, and the Center for Nonlinear and Complex
Systems\\
Duke University, Durham, NC 27708-0129; hsg@cs.duke.edu }

\date{October 20, 1993}

\maketitle

\begin{abstract}

For one-dimensional phase-turbulent solutions of the
Kuramoto-Sivashinsky equation with rigid boundary
conditions, we show that there is a substantial
variation of the correlation time~$\tau_c(x)$ with
spatial position~$x$ in moderately large systems of
size~$L$.  These results suggest that some
time-averaged properties of spatiotemporal chaos do not
become homogeneous away from boundaries for large
systems and for long times.

\end{abstract}


\newpage      % Phys. Letts. requests text begin on new page.

For spatiotemporal chaotic systems (whose spatial
correlation length~$\xi$ is smaller than the system
size~$L$ \cite{Cross93}), an interesting question
arises as to whether time-averaged properties may vary
spatially. In a thermodynamic limit of infinite system
size and for averages over an infinite time interval,
one would expect time-averages to be
spatially-independent as a consequence of ergodicity
and of rotational and translational invariances of the
underlying equations.  For experiments or simulations
of finite-size, however, spatial variation may arise
from boundary conditions, from broken symmetries, or
from a combination of both.

Time-averaged inhomogeneous patterns have indeed been
found for finite-sized systems both numerically and
experimentally. Zaleski \cite{Zaleski85} and Pumir
\cite{Pumir85} studied long-time statistical properties
of spatiotemporal chaotic solutions of the
one-dimensional Kuramoto-Sivashinsky (KS) equation
\begin{equation}
  u_t  +  u_{xx}  +  u_{xxxx}  +  u u_x = 0 ,
  \qquad \mbox{for $0 \le x \le L$} ,
  \label{ks-eq}
\end{equation}
with rigid boundary conditions
\begin{equation}
 u = u_x = 0 . \label{ks-rigid-bcs}
\end{equation}
The time-averaged pattern $p(x) = \langle u(x,t)
\rangle$ \cite{Zaleski85} and simple statistical
quantities like the kurtosis \cite{Pumir85} (derived
from time series measured at a single point in space)
were found to vary spatially, but only for spatial
intervals small compared to the system size and only
near the boundaries. An example is given in
Fig.~\ref{mean-square-averages}, which represents a
more thorough time-averaging than that reported in
Ref.~\cite{Zaleski85}. We found that the maximum
magnitude of the spatial average was independent of the
system size~$L$ over the range $50 \le L \le 2000$,
suggesting that the mean spatial pattern is a
consequence of the boundary conditions.

More recently, Gluckman et al \cite{Gluckman93} found
quite striking time-averaged two-dimensional spatial
patterns in a Faraday crispation experiment for system
sizes that were about five times larger than the
correlation length~$\xi$.  Unlike the results found for
the KS~equation, the time-averaged experimental
patterns showed a highly-ordered periodic lattice with
the symmetry of the lateral boundaries (square or
circular).  Similar results have been reported by Ning
et al for rotating convecting fluid in cells of
different symmetry \cite{Ning93c}.  Although a
systematic study of how the magnitude of these patterns
decreases with aspect ratio was not made, the results
for both numerics and experiments are consistent with
the expected behavior that the magnitude of the
time-averaged patterns should vanish in the
infinite-aspect-ratio limit.

Somewhat related to these investigations of
time-averaged patterns are two studies
\cite{Sato88,Ciliberto87} concerning whether the
fractal dimension~$D_2$ \cite{Eckmann85} varies in
space. This is a possibility since the fractal
dimension is related to the strength and number of
active modes, and these can vary spatially for
finite-aspect-ratio systems \cite{Lorenz91}.  Sato et
al \cite{Sato88} and Ciliberto \cite{Ciliberto87}
obtained their results for chaotic convection in
rectangular cells of modest-sized aspect ratios,
respectively $15 \times 1$ and $4 \times 1$. They
measured long time series at many spatial points and
analyzed these time series with the
Grassberger-Procaccia algorithm \cite{Grassberger83} to
extract~$D_2(x)$.  Sato et al found a strong spatial
dependence of Fourier modes but no clear spatial
dependence of~$D_2$.  Ciliberto found a stronger
spatial dependence of~$D_2$ in the smaller cell. Both
experimental results were not entirely convincing since
dimension algorithms based on time series are difficult
to apply in a reliable way \cite{Ruelle90}.  Although
an unambiguous dependence of fractal dimension on
position was not found, Ciliberto did observe a strong
spatial variation of the root-mean-square of an
observable, with the variation occurring on
lengthscales smaller than a roll size.

We note that the above studies concern issues that are
somewhat different than those raised by the study of
coherent structures in more fully developed turbulent
flows \cite{Cantwell81,Hussain86}. Coherent structures
typically persist only over finite time intervals and
long-time averages have not yet been carried out to see
how the time-averaged pattern may be related to
existing coherent structures.

In this Letter, we look at a different aspect of
spatial-variations of time-averages.  We investigate
the question of whether temporal correlations, as
quantified by a correlation time~$\tau_c$, can vary
spatially. This is a rather different question than
looking at time-averages of fluctuations since time
correlation functions involve comparing information
separated at time intervals that can be quite large. In
the following, we show empirically that there is an
unexpected spatial variation of the correlation time
for a one-dimensional model of spatiotemporal chaos but
only for rigid boundary conditions and only for phase
turbulent states \cite{Kuramoto84}. Unlike the
numerical \cite{Zaleski85,Pumir85} and experimental
\cite{Gluckman93,Ning93c} results discussed above, the
spatial dependence of~$\tau(x)$ is strongest in the
interior, {\em away} from the boundaries. This spatial
dependence exists only in the phase-turbulent KS
equation, is weakly sensitive to the choice of initial
conditions, and disappears with sufficiently long
averaging in the more disordered defect-turbulent
regime of the CGL equation
\cite{Sakaguchi90,Shraiman92}.

One implication of these results is that not all
features of large-aspect-ratio spatiotemporal chaos
become homogeneous away from boundaries, even for
systems that are quite large compared to the
correlation length (for our calculations, this ratio is
about 20).  Experimentalists and computational
scientists will have to be wary of which point they
pick in space to sample information, at least in
systems with long-lived temporal correlations as is the
case for phase turbulence.

We obtained our results by numerical integrations over
times~$T$ as large as~$10^6$ time units for
Eqs.~(\ref{ks-eq}) and~(\ref{ks-rigid-bcs}), and for
the closely related but more general
complex-Ginzburg-Landau (CGL) equation
\cite{Kuramoto84}
\begin{equation}
  \partial_t A(x,t) =
    A
    +  (1 + i c_1) \partial^2_x A
    -  (1 - i c_3) |A|^2 A  ,
  \label{cgl-eq}
\end{equation}
with Dirichlet boundary conditions
\begin{equation}
 A(0,t) = A(L,t) = 0 . \label{cgl-bcs}
\end{equation}
The CGL~field $A(x,t)$ is complex-valued and the
parameters~$c_1$ and~$c_3$ are real-valued.  The KS
equation can be derived \cite{Kuramoto84} from the CGL
equation in the Benjamin-Feir unstable limit~$c_1 c_3
\to 1$.  For fixed~$c_1 \ge 1.8$, there is an
apparent transition from phase turbulence, for which
the field magnitude~$|A|$ is bounded away from zero for
all time, to defect turbulence, in which isolated
space-time defects of zero field~$A$ value occur
\cite{Shraiman92}. It is not yet known whether the two
phases are distinct in the thermodynamic limit of an
infinitely large system, e.g., phase turbulence may
simply be a defect turbulence with an unobservably low
but finite density of space-time defects. Unpublished
results of ours, extending Shraiman et al to much
larger system sizes, suggest that the phases are in
fact distinct.

We integrated Eqs.~(\ref{ks-eq})
and~(\ref{ks-rigid-bcs}) and Eqs.~(\ref{cgl-eq})
and~(\ref{cgl-bcs}) with second-order-accurate
finite-difference codes with time-splitting of the
linear and nonlinear operators \cite{Manneville90}.  At
each of many coordinates~$x$ on a regular spatial mesh
spanning the interval~$[0,L]$, we calculated a
correlation time~$\tau_c(x)$ from the $x$-dependent
time correlation function
\begin{equation}
  C(\tau;x) = \left\langle \,
    \Bigl( u(t+\tau,x)  -  \overline{u}(x) \Bigr) \,
    \Bigl( u(t,x)       -  \overline{u}(x) \Bigr) \,
  \right\rangle_t
  ,  \label{time-correlation-fn}
\end{equation}
where $\overline{u}(x)$ is the function representing
the time averaged mean at each point in space and where
the brackets indicate time averaging over information
at position~$x$.  This function was found to decay
approximately exponentially for small times. We used
the root-mean-square width of~$C(\tau;x)$ up to the
first zero crossing to determine the correlation
time~$\tau$.  Our results (especially
Fig.~\ref{rigid-spatial-variations}) are not sensitive
to our method of extracting~$\tau$, e.g., we found
similar results when estimating~$\tau$ by the
first-zero-crossing of~$C(\tau)$ or by the integral
time scale \cite{Tennekes72}.

We note that the long-time long-wavelength dynamics of
the KS-equation and of the phase-turbulent regime of
the CGL equation is believed to be described by the
Kardar-Parisi-Zhang (KPZ) equation
\cite{Kardar86,Sneppen92,Procaccia92,Grinstein93} which
describes the roughening transition of an interface.
If this asymptotic description is correct, the time
correlation function in one spatial dimension does not
decay exponentially with a single well-defined time
scale, but instead decays as a stretched exponential of
the form $\exp(-B t^{2\beta})$ with~$\beta = 1/3$.  We
ignore this subtlety in what follows since, as shown in
Fig.~\ref{correlation-functions}, there is a
substantial decay of correlations over a finite time
interval and the system sizes are small compared to the
lengthscale over which the KPZ description becomes
valid \cite{Sneppen92}.

We also ignore a possibly serious issue concerning
whether our numerical results are statistically
stationary so that all transients have decayed
sufficiently.  Naively, one would hope that an
integration time~$T$ of order~$10^6$ that is many
orders of magnitude greater than the correlation
time~$\tau_c \le 40$ would be sufficient for transients
to decay.  However, some calculations by Shraiman
\cite{Shraiman86} suggest that KS~phase turbulence with
periodic boundary conditions is transient, with a decay
time that grows exponentially with system size~$L$. (It
is not known if this conclusion generalizes to the
boundary conditions Eq.~(\ref{ks-rigid-bcs}) although
numerical calculations suggest that this is unlikely).
The possibility of extremely long transient times has
also been argued from numerical studies of coupled map
lattices \cite{Crutchfield88,Ershov92}. We do not know
of any quantitative method for distinguishing such
long-lived transients from nontransient chaos, nor do
we know whether there are observable dynamical
implications of averaging over such a transient rather
than over an attractor.  We present the results below
with the hope that, if they are transient, the slow
decay times lead to an ergodic average that is
approximately correct.

Our results are summarized in
Figs.~\ref{correlation-functions}
through~\ref{periodic-spatial-variations}.
Fig.~\ref{correlation-functions} shows parts of the
temporal and spatial correlation functions averaged
over all space and time, from which correlation times
and lengths were calculated for the KS~equation with
rigid boundary conditions. We estimate a value of~$\tau
= 2$ units and of~$\xi = 7$ units respectively and
choose integration times and system sizes much larger
than these values in what follows.

Fig.~\ref{rigid-spatial-variations}(a) shows the
central result of this Letter.  The correlation time of
the KS~equation, as measured by the root-mean-square
width of the temporal correlation function for rigid
boundary conditions, varies substantially with spatial
position. The magnitude of this spatial variation is
quite large, with some peaks being as large as five
times the average background value of about~$8$. The
peaks show structure that varies substantially over
lengthscales comparable to the correlation length.
Fig.~\ref{rigid-spatial-variations}(b) shows how the
temporal correlation function itself varies over a
small spatial interval $25 \le x \le 30$ of the system.
The variation in correlation times between $x=25$ and
$x=30$ is caused by the presence or absence of the
small dip in the correlation function around $\tau =
12$.  The variations in the correlation function at
large $\tau$ remain about the same magnitude out to the
largest values of $\tau$ that we tested.  When averaged
over space, these variations disappear as shown in
Fig.~\ref{correlation-functions}(a).

{}From similar numerical calculations, we have found
empirically that the positions of the peaks in
Fig.~\ref{rigid-spatial-variations}(a) do not change
with the length of the integration time~$T$, with
different initial conditions or for changes in the
numerical time resolution. However, the relative
heights and widths of the peaks do change under these
conditions.

To obtain more insight about the origin of these peaks,
we have made similar calculations for the
one-dimensional complex Ginzburg-Landau equation,
Eq.~(\ref{cgl-eq}), which has a phase-turbulent regime
similar to the KS equation but also a more strongly
disordered defect-turbulent regime for which temporal
correlations decay much more rapidly
\cite{Shraiman92,Egolf93c}.  The spatial variation of
Fig.~\ref{rigid-spatial-variations}(a) is {\em not}
seen for defect-turbulent states with rigid boundary
conditions, Eq.~(\ref{cgl-bcs}), as shown in
Fig.~\ref{cgl-rigid-variations}.
Fig.~\ref{cgl-rigid-variations}(a) shows the time
averaged mean of the amplitude of the field $A$ as a
function of position while
Fig.~\ref{cgl-rigid-variations}(b) shows how the
correlation time varies with position.  In both cases,
peaks are observed near the boundaries while the bulk
variations are small and smooth.  The asymmetry in the
heights of the outermost large peaks may be a
consequence of some small initial asymmetry in the
initial conditions. (We have not studied rigid boundary
condition solutions in the phase-turbulent regime of
the CGL~equation, which presumably would give results
similar to Fig.~\ref{rigid-spatial-variations}.) We
deduce that the spatial variation of the correlation
time is a consequence of the long-lived time
correlations present in phase turbulence.

We have also tested the importance of rigid boundary
conditions by numerically integrating the 1d
CGL~equation with periodic boundary conditions in the
phase-turbulent regime. In this case, we find that
there is no spatial variation of the correlation time
as shown in Fig.~\ref{periodic-spatial-variations}.
Essentially identical results are found for the KS
equation with periodic boundary conditions with the
difference that one finds substantial spatial
variations for short integration times that slowly
decay away in magnitude for longer integration times.
These results confirm our earlier summary of previous
numerical and laboratory experiments
\cite{Zaleski85,Pumir85,Gluckman93,Ning93c}, that
average spatial patterns are driven by the presence of
boundary conditions that break the translational
symmetry.

In conclusion, we have looked at a variation of the
idea that time-average spatiotemporal chaotic patterns
can have interesting mean structure. By numerical
simulations on the KS and CGL equations in one-space
dimension, we have found that the correlation
time~$\tau_c$ can vary substantially in space when
there are long-lived temporal correlations (phase
turbulence) and for boundary conditions that break
translational invariance. For dynamical states with
more rapidly decaying correlation functions (defect
turbulence) or for periodic boundary conditions, there
is no interesting spatial variation of~$\tau_c$
provided that one integrates sufficiently long in time.
It would be interesting to explore these issue further
in laboratory experiments on fluids and flames.

We thank Geoff Grinstein for several useful
discussions.  This work was supported by National
Science Foundation Grant ASC-8820327, by a fellowship
from the Office of Naval Research, and by allotments of
CRAY CPU~time at the North Carolina Supercomputing
Center and at the National Center for Supercomputing
Applications.

%\bibliographystyle{prsty}  % entries in order of citation
%\bibliography{bibliography,hsg}

\begin{thebibliography}{10}

\bibitem{Cross93}
M.~C. Cross and P.~C. Hohenberg, Rev. Mod. Phys. {\bf 65},  851  (1993).

\bibitem{Zaleski85}
S. Zaleski and P. Lallemand, J. Physique Lett. {\bf 46},  L793  (1985).

\bibitem{Pumir85}
A. Pumir, J. Physique {\bf 46},  511  (1985).

\bibitem{Gluckman93}
B.~J. Gluckman, P. Marcq, J. Bridger, and J.~P. Gollub, Phys. Rev. Lett. {\bf
  71},  2034  (1993).

\bibitem{Ning93c}
L. Ning, Y. Hu, R.~E. Ecke, and G. Ahlers, Phys. Rev. Lett. {\bf 71},  2216
  (1993).

\bibitem{Sato88}
S. Sato, M. Sano, and Y. Sawada, Phys. Rev. A {\bf 37},  1679  (1988).

\bibitem{Ciliberto87}
S. Ciliberto, Europhys. Lett. {\bf 4},  685  (1987).

\bibitem{Eckmann85}
J.-P. Eckmann and D. Ruelle, Rev. Mod. Phys. {\bf 57},  617  (1985).

\bibitem{Lorenz91}
E.~N. Lorenz, Nature {\bf 353},  241  (1991).

\bibitem{Grassberger83}
P. Grassberger and I. Procaccia, Physica {\bf D9},  189  (1983).

\bibitem{Ruelle90}
D. Ruelle, Proc. R. Soc. London {\bf A427},  241  (1990).

\bibitem{Cantwell81}
B.~J. Cantwell, Ann. Rev. Fluid Mech. {\bf 13},  457  (1981).

\bibitem{Hussain86}
A.~K. M.~F. Hussain, J. Fluid. Mech {\bf 173},  303  (1986).

\bibitem{Kuramoto84}
Y. Kuramoto, {\em Chemical Oscillations, Waves, and Turbulence}
  (Springer-Verlag, New York, 1984).

\bibitem{Sakaguchi90}
H. Sakaguchi, Prog. Theor. Phys. {\bf 84},  792  (1990).

\bibitem{Shraiman92}
B.~I. Shraiman {\it et~al.}, Physica D {\bf 57},  241  (1992).

\bibitem{Manneville90}
P. Manneville, {\em Dissipative Structures and Weak Turbulence}, {\em
  Perspectives In Physics} (Academic Press, New York, 1990).

\bibitem{Tennekes72}
H. Tennekes and J.~L. Lumley, {\em A First Course in Turbulence} (The MIT
  Press, Cambridge, Mass., 1972).

\bibitem{Kardar86}
M. Kardar, G. Parisi, and Y.-C. Zhang, Phys. Rev. Lett. {\bf 56},  889  (1986).

\bibitem{Sneppen92}
K. Sneppen {\it et~al.}, Phys. Rev. A {\bf 46},  7351  (1992).

\bibitem{Procaccia92}
I. Procaccia {\it et~al.}, Phys. Rev. A {\bf 46},  3230  (1992).

\bibitem{Grinstein93}
G. Grinstein and C. Jayaprakash (unpublished).

\bibitem{Shraiman86}
B.~I. Shraiman, Phys. Rev. Lett. {\bf 57},  325  (1986).

\bibitem{Crutchfield88}
J.~P. Crutchfield and K. Kaneko, Phys. Rev. Lett. {\bf 60},  2715  (1988).

\bibitem{Ershov92}
S.~V. Ershov and A.~B. Potapov, Phys. Letts. A {\bf 167},  60  (1992).

\bibitem{Egolf93c}
D.~A. Egolf and H.~S. Greenside, submitted to Nature. Available as a preprint
  from the Los Alamos archive, number patt-sol/9307010 (unpublished).

\end{thebibliography}

\newpage
\begin{figure}   % first figure
\caption{
Spatial dependence of {\bf (a)} the time-averaged
mean-value $\langle u \rangle$ and of {\bf (b)} the
time-averaged square-deviation $\langle (u - \langle u
\rangle)^2 \rangle$ for solutions of the
one-dimensional Kuramoto-Sivashinsky equation
Eq.~(\protect\ref{ks-eq}) with rigid boundary
conditions Eq.~(\protect\ref{ks-rigid-bcs}) on an
interval of size~$L=100$. The temporal resolution
was~$\triangle{t} = 0.1$, the spatial resolution
was~$\triangle{x} = 0.5$, and the total integration
time was~$T=10^6$ time units. For initial data, we used
uniformly distributed random numbers in the
interval~$[-0.03,0.03]$. }
\label{mean-square-averages}
\end{figure}

\begin{figure}   % second figure
\caption{
{\bf (a)} Plot of the temporal correlation function
Eq.~(\protect\ref{time-correlation-fn}) averaged over
all space and time for a numerical solution of the
KS~equation, Eqs.~(\protect\ref{ks-eq})
and~(\protect\ref{ks-rigid-bcs}).  There is a
substantial decay of correlation over a time
scale~$\tau_c = 2$. The numerical parameters are the
same as those given in
Fig.~\protect\ref{mean-square-averages}.  {\bf (b)}
Space- and time-averaged spatial correlation
function~$S(y) = \left\langle \, \Bigl( u(t,x+y) -
\overline{u}(t) \Bigr) \, \Bigl( u(t,x) -
\overline{u}(t) \Bigr) \, \right\rangle_{x, t} $
plotted over the interval $0 \le x \le 50$ for a system
of size~$L=100$. There is an oscillatory exponential
decay of correlations, defining a spatial correlation
length~$\xi = 7$ that is small compared to the system
size~$L$.  }
\label{correlation-functions}
\end{figure}

\begin{figure}   % third figure
\caption{
{\bf (a)} Plot of the root-mean-square (rms) width of
the correlation function~$C(\tau)$ as a function of
position~$x$, for the same parameters used in
Figs~\protect\ref{mean-square-averages}
and~\protect\ref{correlation-functions}. A substantial
spatial dependence of~$\tau_c$ is found. {\bf (b)}
Details of the variations of~$C(\tau)$ over the spatial
interval~$25 \le x \le 30$.  The initial decay of the
correlation function has not been plotted to allow
smaller features of the function to be seen on the same
scale. }
\label{rigid-spatial-variations}
\end{figure}

\begin{figure}   % fourth figure
\caption{
{\bf (a)} The time-averaged mean value and {\bf (b)}
the rms width of the time correlation
function~$C(\tau)$ of the amplitude of the field in the
CGL~equation with rigid boundary conditions,
Eqs.~(\protect\ref{cgl-eq})
and~(\protect\ref{cgl-bcs}). Results are presented for
the parameter values~$c_1=3.5$ and~$c_3=1.5$, which is
in the defect-turbulent regime \protect\cite{Shraiman92}.
Spatial variations are found only near the boundaries.
The temporal resolution was $\triangle{t} = 0.05$, the
spatial resolution was $\triangle{x} = 0.25$, and the
total integration time was $T=4 \times 10^6$ time
units.}
\label{cgl-rigid-variations}
\end{figure}

\begin{figure}   % fifth figure
\caption{
Plot of the rms width of the time correlation
function~$C(\tau)$ for periodic boundary conditions of
the 1d CGL equation for parameter values $(c1, c3) =
(3.5, 0.7)$. The numerical parameters were~$L=100$,
$256$ Fourier modes, $\triangle{t} = 0.2$, and~$T=10^6$
time units. After a sufficiently long time-averaging,
the correlation time is no longer varying spatially. }
\label{periodic-spatial-variations}
\end{figure}

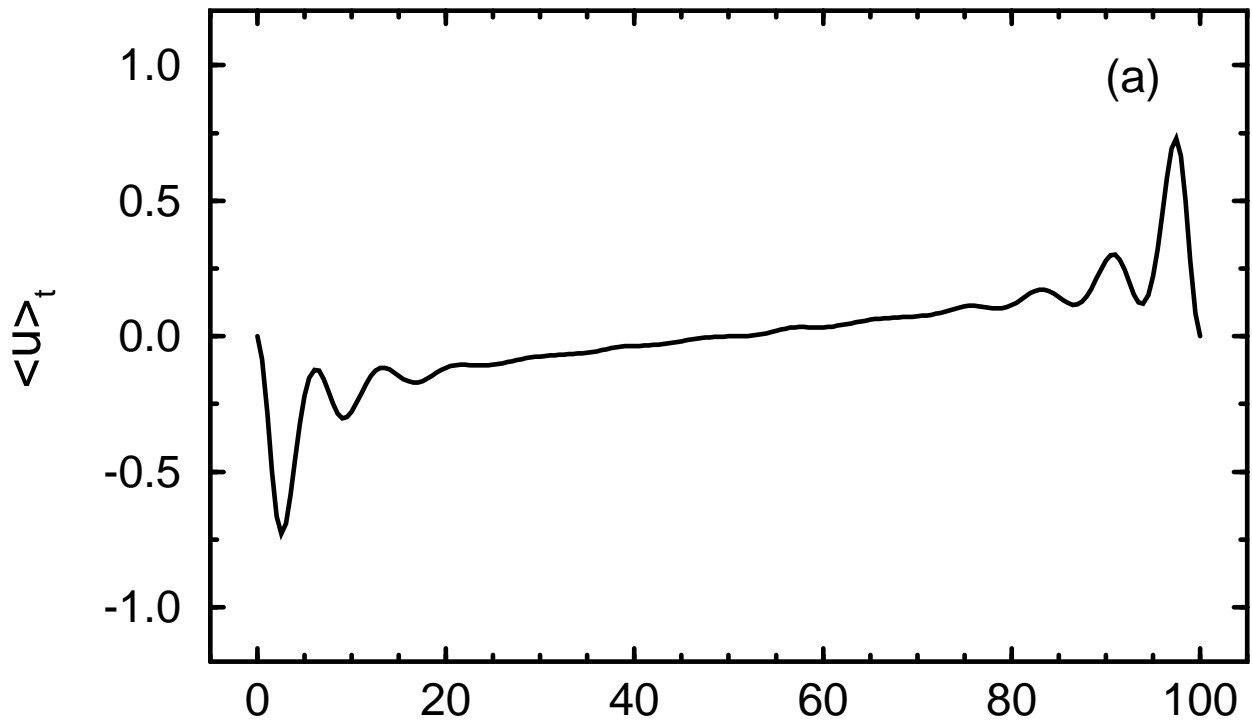
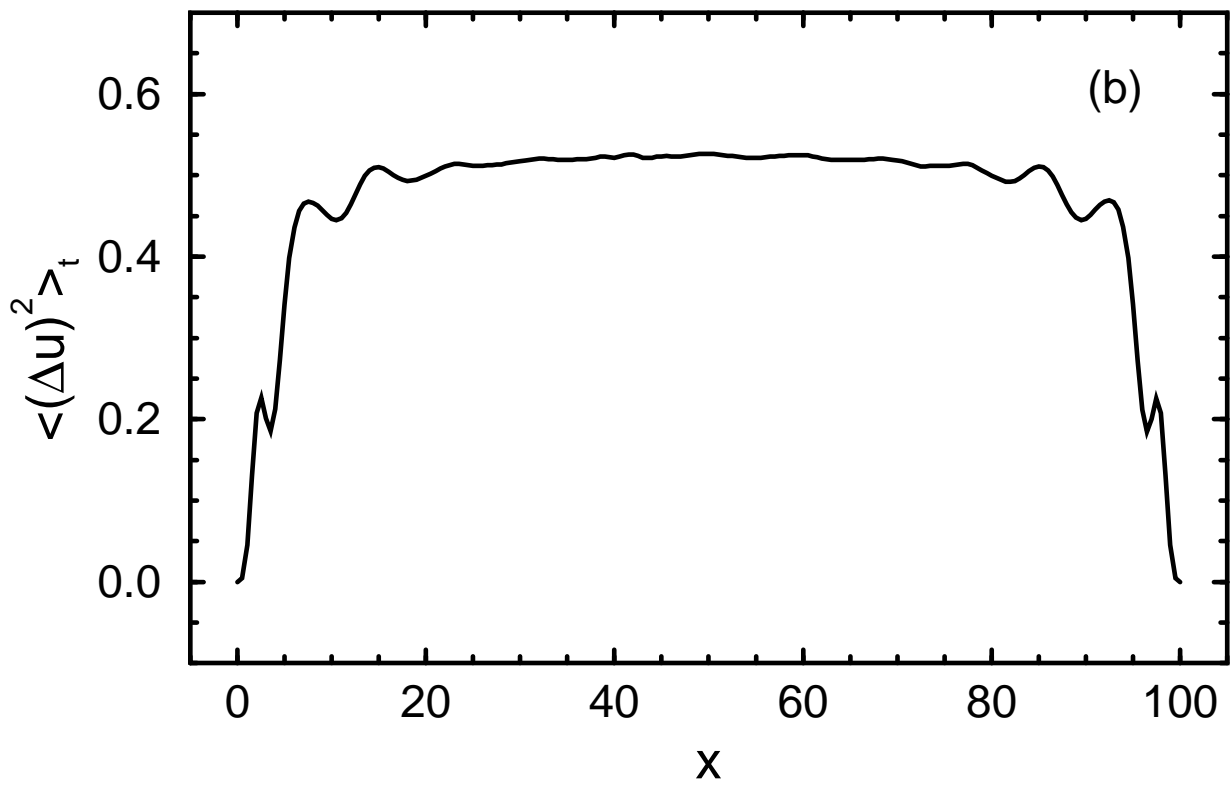
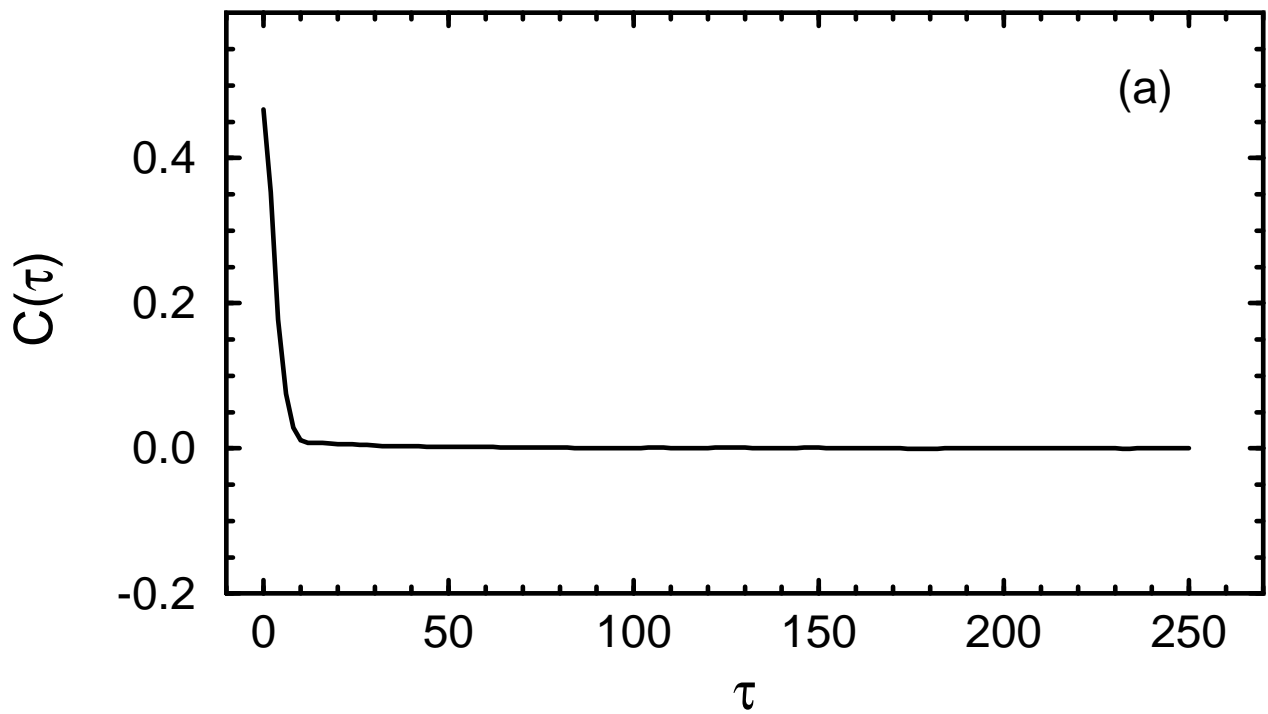
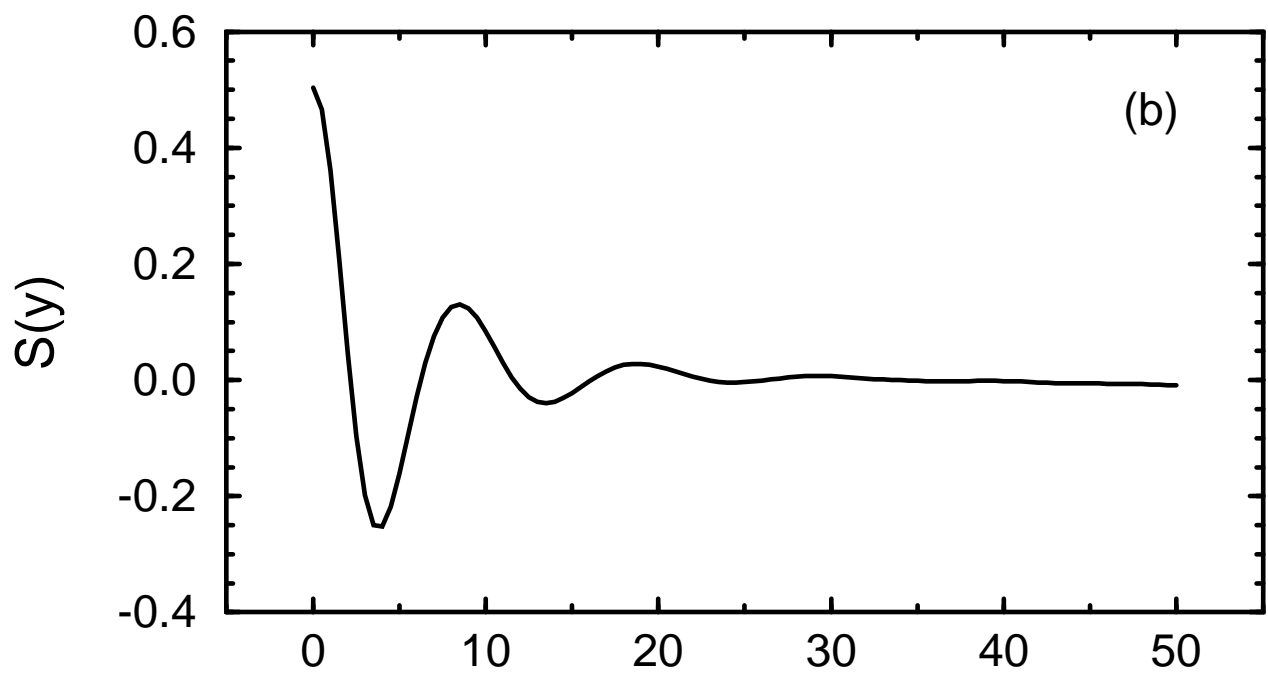
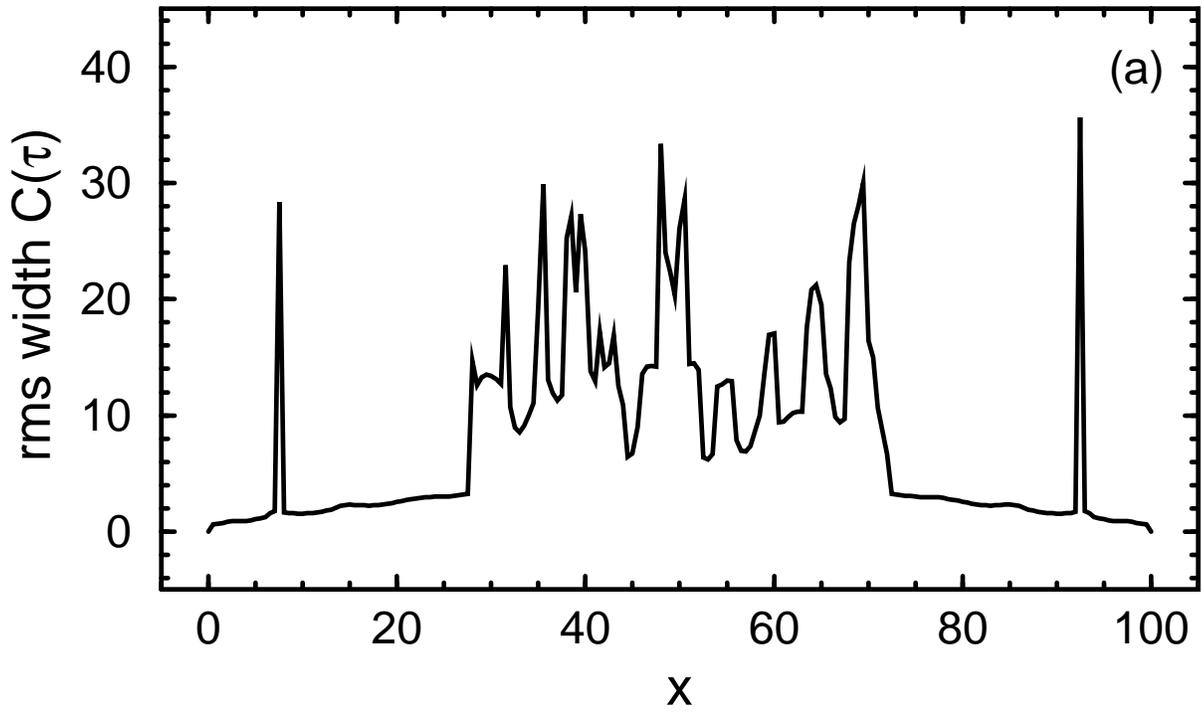
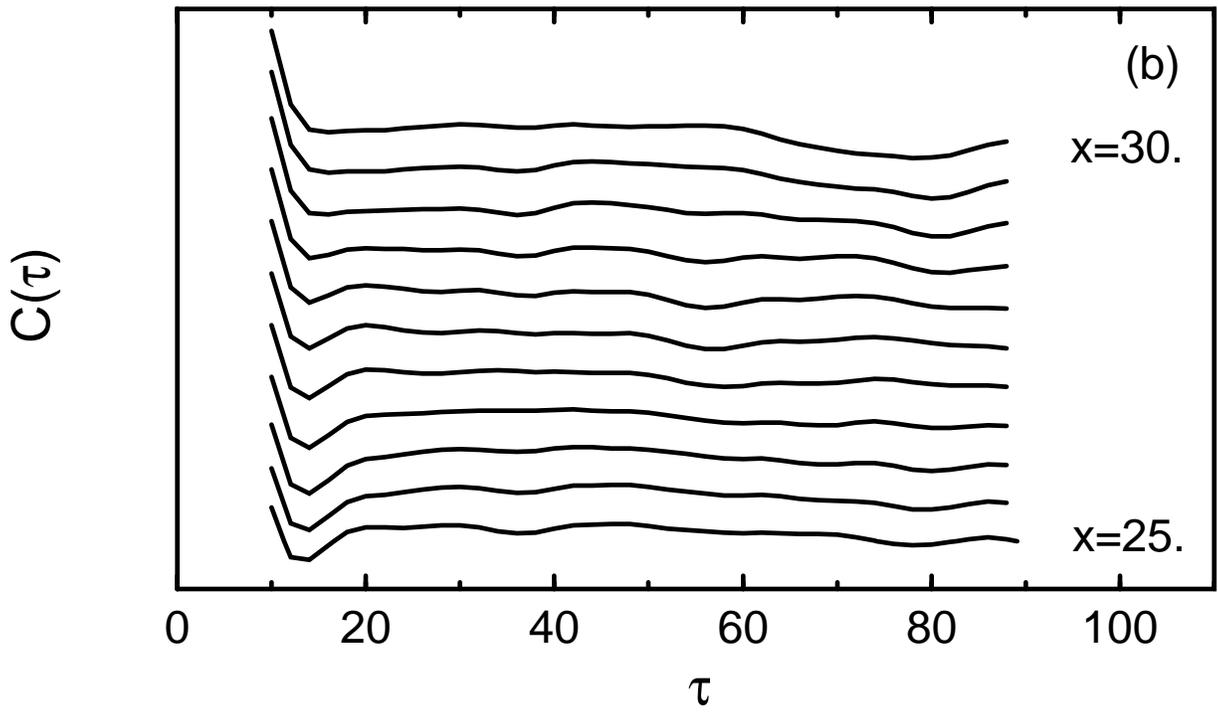
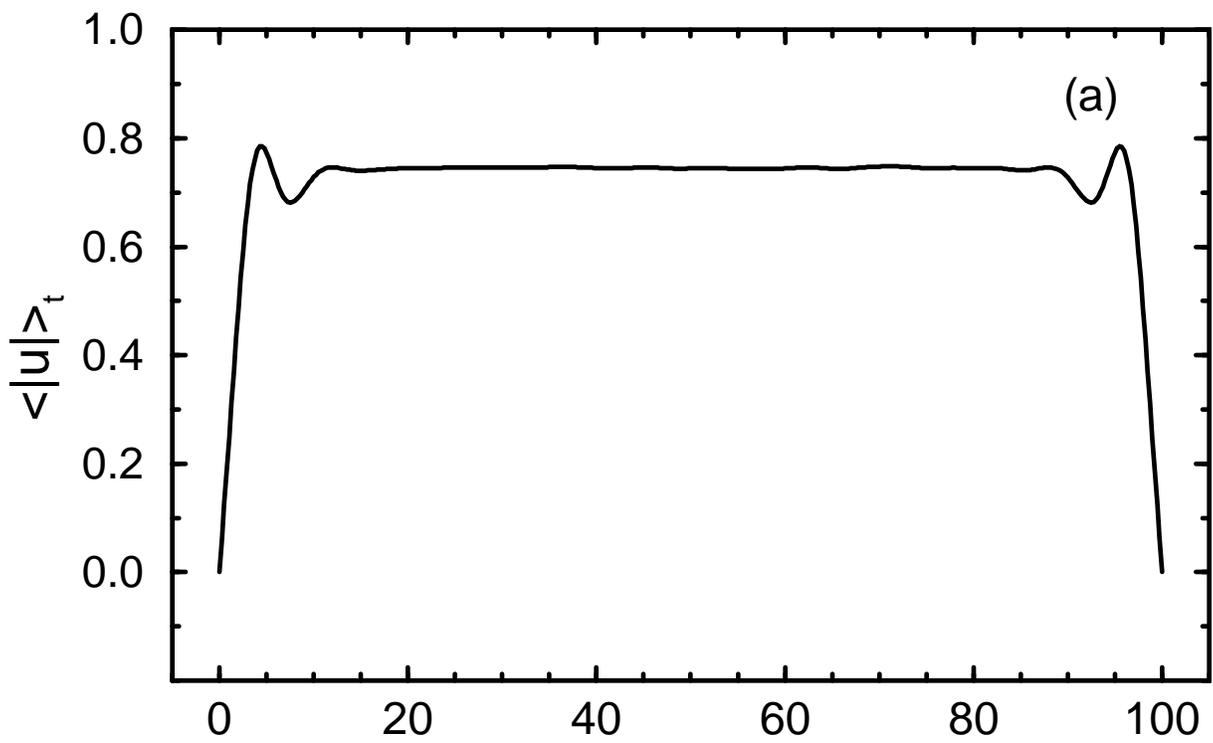
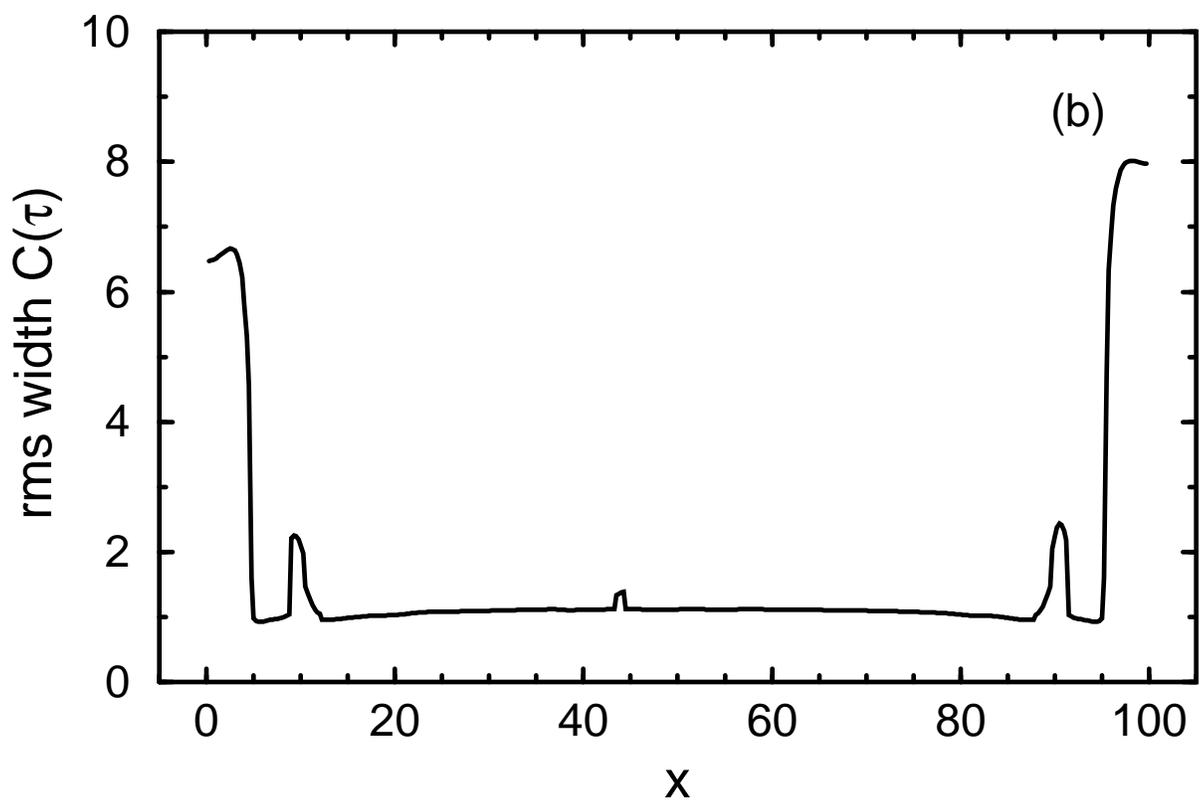
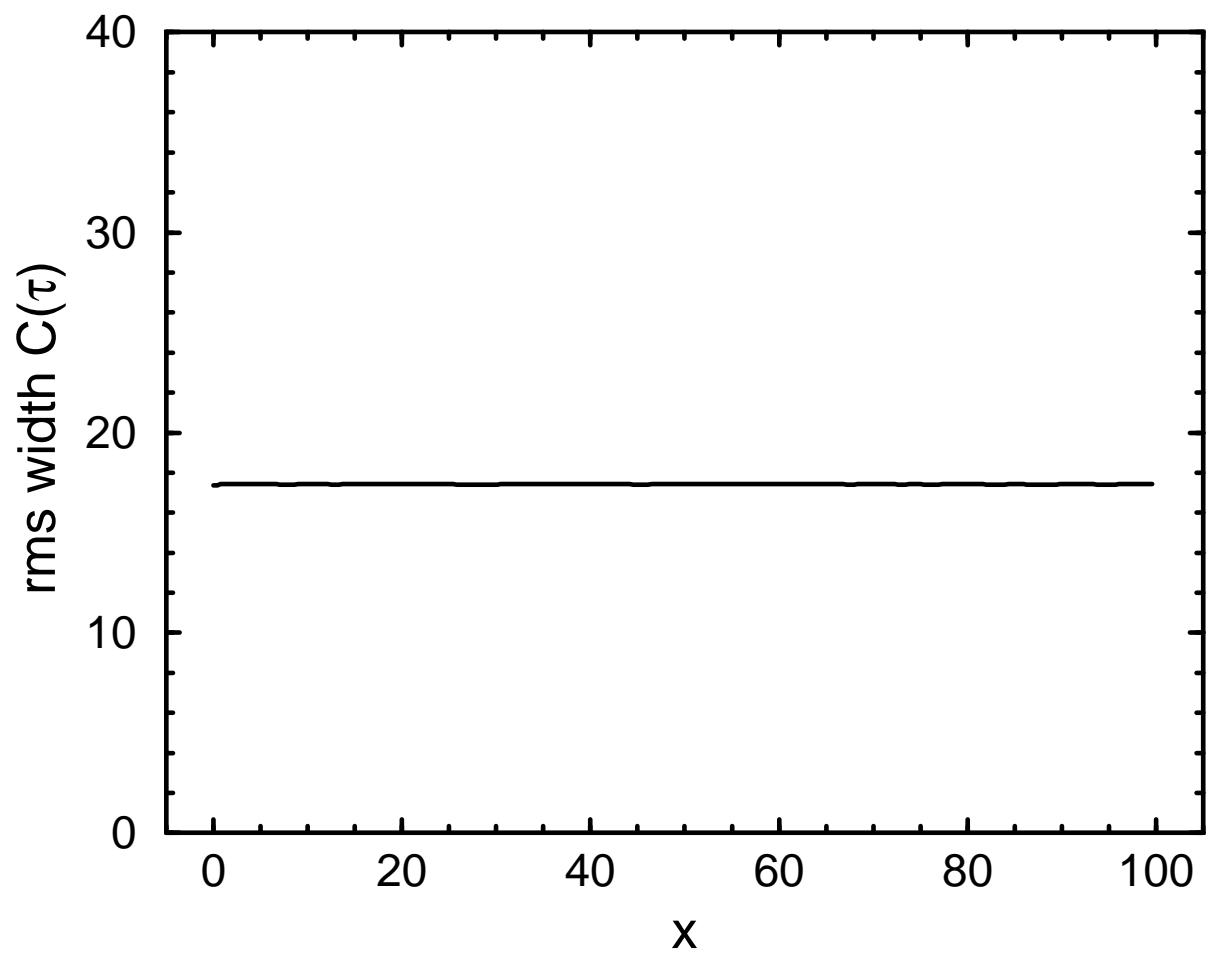
\end{document}